\documentclass{PoS}
\usepackage{wrapfig}
\usepackage{epsfig}
\def \et {E_{T}}
\newcommand{\met}{\mbox{$\not\!\!\et$}}
\newcommand{\aZerow}{a_0^W}

\newcommand{\aCw}{a_C^W}
\newcommand{\aCz}{a_C^Z}

\title{Anomalous quartic and triple gauge 
couplings in $\gamma$-induced processes at the LHC}

\ShortTitle{Anomalous quartic and triple gauge couplings}

\author{\speaker{Christophe Royon}\\
        CEA/IRFU/Service de physique des particules, CEA/Saclay, 91191 
Gif-sur-Yvette cedex, France\\
        E-mail: \email{christophe.royon@cea.fr}}

\author{Emilien Chapon\\
        CEA/IRFU/Service de physique des particules, CEA/Saclay, 91191 
Gif-sur-Yvette cedex, France\\
        E-mail: \email{emilien.chapon@cea.fr}}

\author{Old\v{r}ich Kepka\\
        IPNP, Faculty of Mathematics and Physics,
Charles University, Prague\\
Center for Particle Physics, Institute of Physics, Academy of Science, Prague\\
        E-mail: \email{kepkao@fzu.cz}}

\abstract{We study the $W/Z$ pair production via two-photon exchange at the LHC and
give the sensitivities on trilinear and quartic gauge anomalous couplings 
between photons and $W/Z$ bosons for an integrated luminosity of 30 and 
200 fb$^{-1}$. For simplicity and to obtain lower backgrounds, only the leptonic
decays of the electroweak bosons are considered. The intact protons in the final
states are detected in the ATLAS Forward Proton detectors. The high energy and
luminosity of the LHC and the forward detectors allow to probe beyond standard model physics and to test the
higgsless and extra dimension models in an unprecedent way.
}

\FullConference{XVIII International Workshop on Deep-Inelastic Scattering and Related Subjects, DIS 2010\\
		April 19-23, 2010\\
		Firenze, Italy}

\begin{document}
In the Standard Model (SM) of particle physics, the couplings of fermions and 
gauge bosons are constrained by the gauge symmetries of the Lagrangian.
The measurement of $W$ and $Z$ boson pair productions via the exchange of
two photons  
allows to provide directly stringent tests
of one of the most important and least understood
mechanism in particle physics, namely the
electroweak symmetry breaking~\cite{stirling}. The non-abelian gauge nature of the SM
predicts the existence of quartic couplings
$WW\gamma \gamma$
between the $W$ bosons and the photons which can be probed directly at the 
Large Hadron Collider (LHC) at CERN.
The quartic coupling to the $Z$ boson $ZZ\gamma \gamma$ is not present in the
SM. 

The quartic couplings test 
more generally new physics which couples to electroweak bosons.
Exchange of heavy particles beyond the SM might manifest itself as a
modification of the quartic couplings appearing in contact 
interactions~\cite{higgsless}. It is
also worth noticing that in the limit of infinite Higgs masses, or in Higgs-less
models~\cite{higgsless}, new structures not present in the tree level Lagrangian
appear in the quartic $W$ coupling. For
example, if the electroweak breaking mechanism does not manifest itself in the
discovery of the Higgs boson at the LHC or supersymmetry, the presence of
anomalous, non SM like, couplings might be the first evidence of new
physics in the electroweak sector of the SM.

\section{Photon exchange processes in the SM}
The process that we intend to study is the $W$ pair production shown in Fig.~1
induced by the 
exchange of two photons~\cite{piotr,us}. It is a pure QED process
in which the decay products of the $W$ bosons are measured in the central 
detector and the scattered protons leave intact in
the beam pipe at very small angles, contrary to inelastic collisions. Since 
there is no proton remnant the process is purely exclusive; only $W$ decay products 
populate the central detector, and the intact protons can be detected in
dedicated detectors located along the beam line far away from the interaction
point.

The cross section of the $pp\rightarrow p WW p$ process which proceeds through 
two-photon exchange is calculated as a convolution of the 
two-photon luminosity and the total cross section $\gamma\gamma\rightarrow WW$.
The total two-photon cross section is 95.6 fb. 

All considered processes (signal and background) were produced using the Forward
Physics Monte Carlo~\cite{fpmc} (FPMC) generator. The aim of FPMC is to produce different
kinds of processes such as inclusive and exclusive diffraction, photon-exchange
processes. FPMC was interfaced to as fast simulation of the ATLAS
detector~\cite{atlfast}. To reduce the amount of considered background, we only
use leptonic (electrons and muons) decays of $Z$ and $W$ bosons. The clean two-leptonic signature of the two boson signal process
$\gamma\gamma\rightarrow W^{+}W^{-} \rightarrow l\bar{l}\nu\bar{\nu} $
can be mimicked by several background processes which all have
two intact protons in the final state. They are the following:
\begin{enumerate}
\item $\gamma\gamma\rightarrow l\bar{l}$ - two-photon dilepton production 
\item DPE$\rightarrow l\bar{l}$ - dilepton production through double pomeron 
exchange
\item DPE$\rightarrow W^+W^-\rightarrow l\bar{l}\nu\bar{\nu}$ - diboson 
production through double pomeron exchange
\end{enumerate}

After simple cuts to select exclusive $W$ pairs decaying into leptons, such
as a cut on the proton momentum loss of the proton ($0.0015<\xi<0.15$) --- we
assume the protons to be tagged in the ATLAS Forward Physics
detectors~\cite{afp} ---
on the transverse momentum of the leading and second leading leptons at 25 and
10 GeV respectively, on $\met>20$ GeV, $\Delta \phi>2.7$ between leading
leptons, and $160<W<500$ GeV, the diffractive mass reconstructed using the
forward detectors, the background is found to be less than 1.7 event for 30
fb$^{-1}$ for a SM signal of 51 events. In this channel, a 5 $\sigma$ discovery
of the Standard Model $pp\rightarrow pWWp$ process is possible after 5 fb$^{-1}$.  

\begin{figure}[t]
\hfill
\begin{minipage}[t]{.45\textwidth}

\epsfig{file=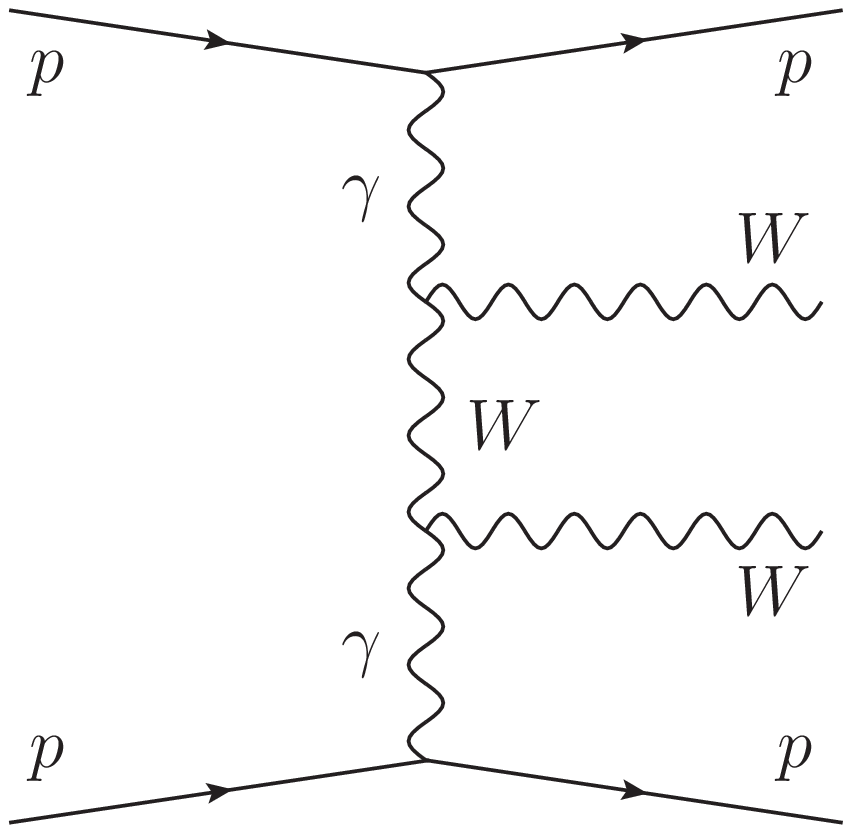,width=6.cm} 
\caption{Sketch diagram showing the two-photon production of a central system.}

\end{minipage}
\hfill
\begin{minipage}[t]{.45\textwidth}

\epsfig{file=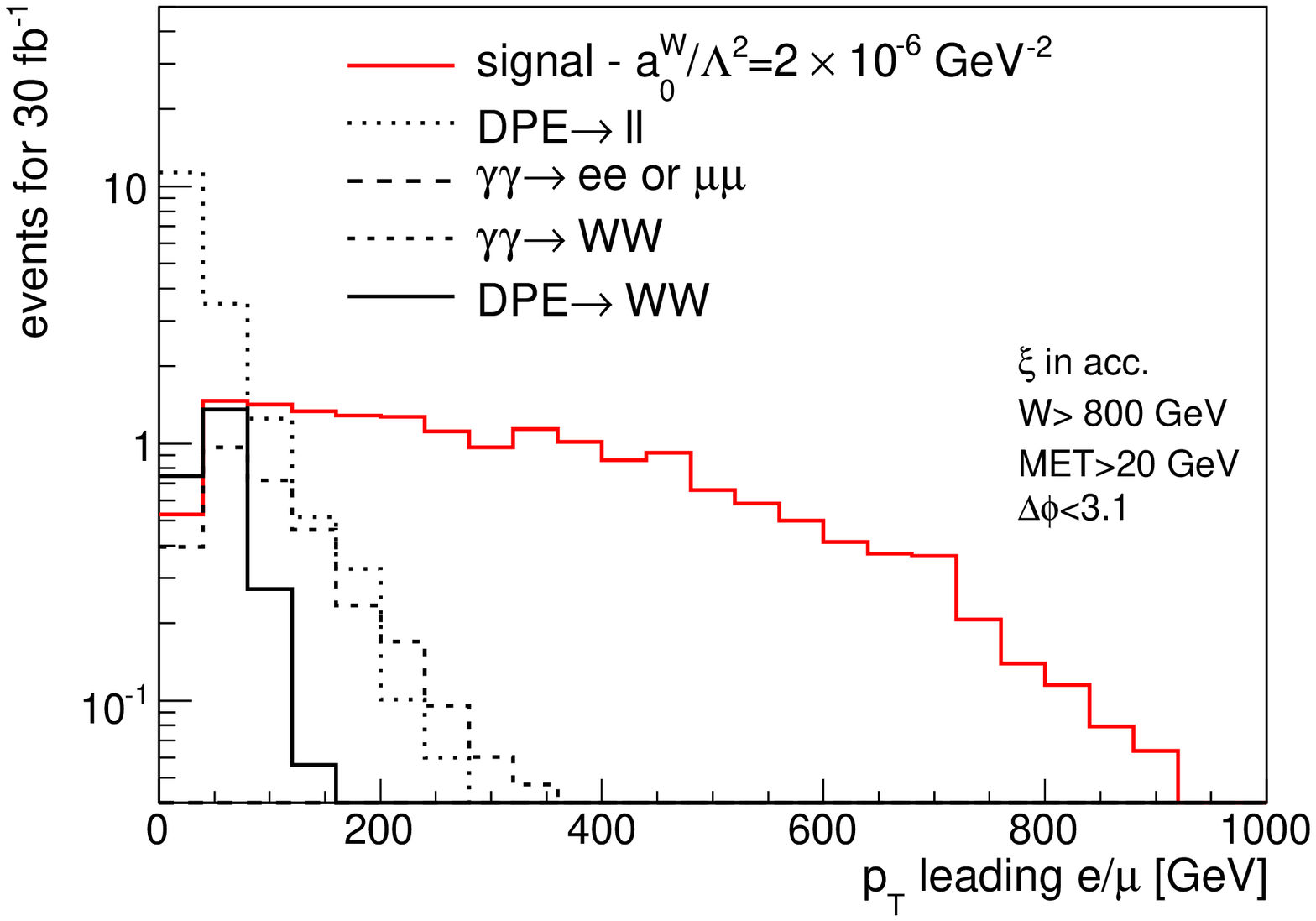,width=6.cm} 
\caption{Distribution of the transverse momentum of the leading lepton for
signal and background after the cut on $W$, $\met$, and $\Delta \phi$ between
the two leptons.}

\end{minipage}
\hfill
\end{figure}

\section{Quartic anomalous couplings}
The parameterization of the quartic couplings
based on \cite{Belanger:1992qh} is adopted. We concentrate on the lowest order 
dimension operators which have
the correct Lorentz invariant structure and obey the $SU(2)_C$ 
custodial symmetry in order to fulfill the stringent experimental bound on the 
$\rho$ parameter. The lowest order interaction
Lagrangians which involve two photons are dim-6 operators. 
The following expression for the effective quartic Lagrangian is used
 \begin{eqnarray}
     \mathcal{L}_6^0 &=& \frac{-e^2}{8} \frac{\aZerow}{\Lambda^2} F_{\mu\nu} 
     F^{\mu\nu} W^{+\alpha} W^-_\alpha 
     - \frac{e^2}{16\cos^2 \theta_W} \frac{a^Z_0}{\Lambda^2} F_{\mu\nu} 
     F^{\mu\nu} Z^\alpha
Z_\alpha\nonumber \\
     \mathcal{L}_6^C & = & \frac{-e^2}{16} \frac{\aCw}{\Lambda^2} 
     F_{\mu\alpha} F^{\mu\beta} (W^{+\alpha} W^-_\beta + W^{-\alpha} 
     W^+_\beta) 
	- \frac{e^2}{16\cos^2 \theta_W} \frac{a^Z_C}{\Lambda^2} 
	F_{\mu\alpha} F^{\mu\beta} Z^\alpha Z_\beta
\label{eq:anom:lagrqgc}
\end{eqnarray}
where $a_0$, $a_C$ are the parametrized new coupling constants and the new scale $\Lambda$ is introduced
so that the Lagrangian density has the correct dimension four and is 
interpreted as the typical mass scale of  new
physics.
In the above formula, we allowed the $W$ and $Z$ parts of the Lagrangian to 
have specific couplings, i.e.
 $a_0\rightarrow (\aZerow$, $a^Z_0$) and similarly $a_C\rightarrow(\aCw$, 
 $\aCz$).

The $WW$ and $ZZ$ two-photon cross sections rise quickly at high energies when 
any of the anomalous parameters are non-zero. The cross section rise has to be 
regulated by a form factor which vanishes in the high energy limit to 
construct a realistic physical model of the BSM theory. We therefore 
modify the couplings by form factors 
that have the desired behavior, i.e. they modify the coupling at small 
energies only slightly but suppress it  when the center-of-mass energy $W_{\gamma\gamma}$ 
increases. The form of the form factor that we consider is the following
\begin{eqnarray}
a\rightarrow \frac{a}{(1+W^2_{\gamma\gamma}/\Lambda^2)^n}
\label{eq:anom:formfactor}
\end{eqnarray}
where $n$=2, and $\Lambda \sim$2 TeV.

The cuts to select quartic anomalous gauge coupling $WW$ events are similar as the
ones we mentioned in the previous section, namely $0.0015<\xi<0.15$ for the
tagged protons, $\met>$ 20 GeV, $\Delta \phi<3.13$ between the two leptons. In
addition, a cut on the $p_T$ of the leading lepton $p_T>160$ GeV and on the
diffractive mass $W>800$ GeV are requested since anomalous coupling events
appear at high mass. Fig~2 displays the $p_T$ distribution of the leading lepton
for signal and the different considered backgrounds. 
After these requirements, we expect about 0.7 background
events for an expected signal of 17 events if the anomalous coupling is about
four order of magnitude lower than the present LEP limit ($|a_0^W / \Lambda^2| =
5.4$ 10$^{-6}$) for a luminosity of 30 fb$^{-1}$. The strategy to select anomalous coupling $ZZ$ events is
analogous and the presence of three leptons or two like sign leptons are 
requested. The expected number of events for a luminosity of 30 fb$^{-1}$ as a
function of the anomalous coupling is displayed in Fig~3 and the $5\sigma$
discovery countours for $W$ and $Z$ quartic anomalous couplings are displayed
for two different integrated luminosities of 30 and 200 fb$^{-1}$ in Fig.~4.
Table 1 gives the reach on anomalous couplings at the LHC for the same
luminosities compared to the present OPAL limits~\cite{opal}. We note that we can gain almost
four orders of magnitude in the sensitivity to anomalous quartic gauge couplings
compared to LEP experiments, and it is possible to reach the values expected in Higgsless
or extra-dimension models. The tagging of the protons using the ATLAS Forward
Physics detectors is the only method at present to test so small values of
quartic anomalous couplings and thus to probe the higgsless models in a clean
way.

\begin{table}
\begin{center}
   \begin{tabular}{|c||c|c|c|}
    \hline
    Couplings & 
    OPAL limits & 
    \multicolumn{2}{c|}{Sensitivity @ $\mathcal{L} = 30$ (200) fb$^{-1}$} \\
    &  \small[GeV$^{-2}$] & 5$\sigma$ & 95\% CL \\ 
    \hline
    $a_0^W/\Lambda^2$ & [-0.020, 0.020] & 5.4 10$^{-6}$ & 2.6 10$^{-6}$\\
                      &                 & (2.7 10$^{-6}$) & (1.4 10$^{-6}$)\\ \hline               
    $a_C^W/\Lambda^2$ & [-0.052, 0.037] & 2.0 10$^{-5}$ & 9.4 10$^{-6}$\\
                      &                 & (9.6 10$^{-6}$) & (5.2 10$^{-6}$)\\ \hline               
    $a_0^Z/\Lambda^2$ & [-0.007, 0.023] & 1.4 10$^{-5}$ & 6.4 10$^{-6}$\\
                      &                 & (5.5 10$^{-6}$) & (2.5 10$^{-6}$)\\ \hline               
    $a_C^Z/\Lambda^2$ & [-0.029, 0.029] & 5.2 10$^{-5}$ & 2.4 10$^{-5}$\\
                      &                 & (2.0 10$^{-5}$) & (9.2 10$^{-6}$)\\ \hline               
    \hline
   \end{tabular}
\end{center}
\caption{Reach on anomalous couplings obtained in $\gamma$ induced processes
after tagging the protons in the final state in the ATLAS Forward Physics
detectors compared to the present OPAL limits. The $5\sigma$ discovery and 95\%
C.L. limits are given for a luminosity of 30 and 200 fb$^{-1}$} 
\end{table}

\section{Triple anomalous gauge couplings}
In Ref.~\cite{Kepka:2008yx}, we also studied the sensitivity to triple gauge
anomalous couplings at the LHC. The Lagrangian including anomalous triple gauge 
couplings $\lambda^{\gamma}$ and $\Delta\kappa^{\gamma}$ is the following
\begin{eqnarray}
   \mathcal{L} &\sim& (W^{\dagger}_{\mu\nu}W^{\mu}A^{\nu}-W_{\mu\nu}W^{\dagger\mu}A^{\nu})
   \nonumber \\
   &~&
+(1+\Delta\kappa^{\gamma})W_{\mu}^{\dagger}W_{\nu}A^{\mu\nu}+\frac{{\lambda^{\gamma}}}{M_W^2}W^{\dagger}_{\rho\mu}
   W^{\mu}_{\phantom{\mu}\nu}A^{\nu\rho} ).
\end{eqnarray}
The strategy is the same as for quartic couplings: we first implement this
lagragngian in FPMC and we select the signal events when the $Z$ and $W$ bosons
decay into leptons. The difference is that the signal appears at high mass for 
$\lambda^{\gamma}$ and $\Delta
\kappa^{\gamma}$ only modifies the normalisation and the low mass events have to
be retained. The sensitivity on triple gauge anomalous couplings is a gain of
about a factor 3 with respect to the LEP limits, which represents one of the
best reaches before the LHC.

\begin{figure}[t]
\hfill
\begin{minipage}[t]{.45\textwidth}

\epsfig{file=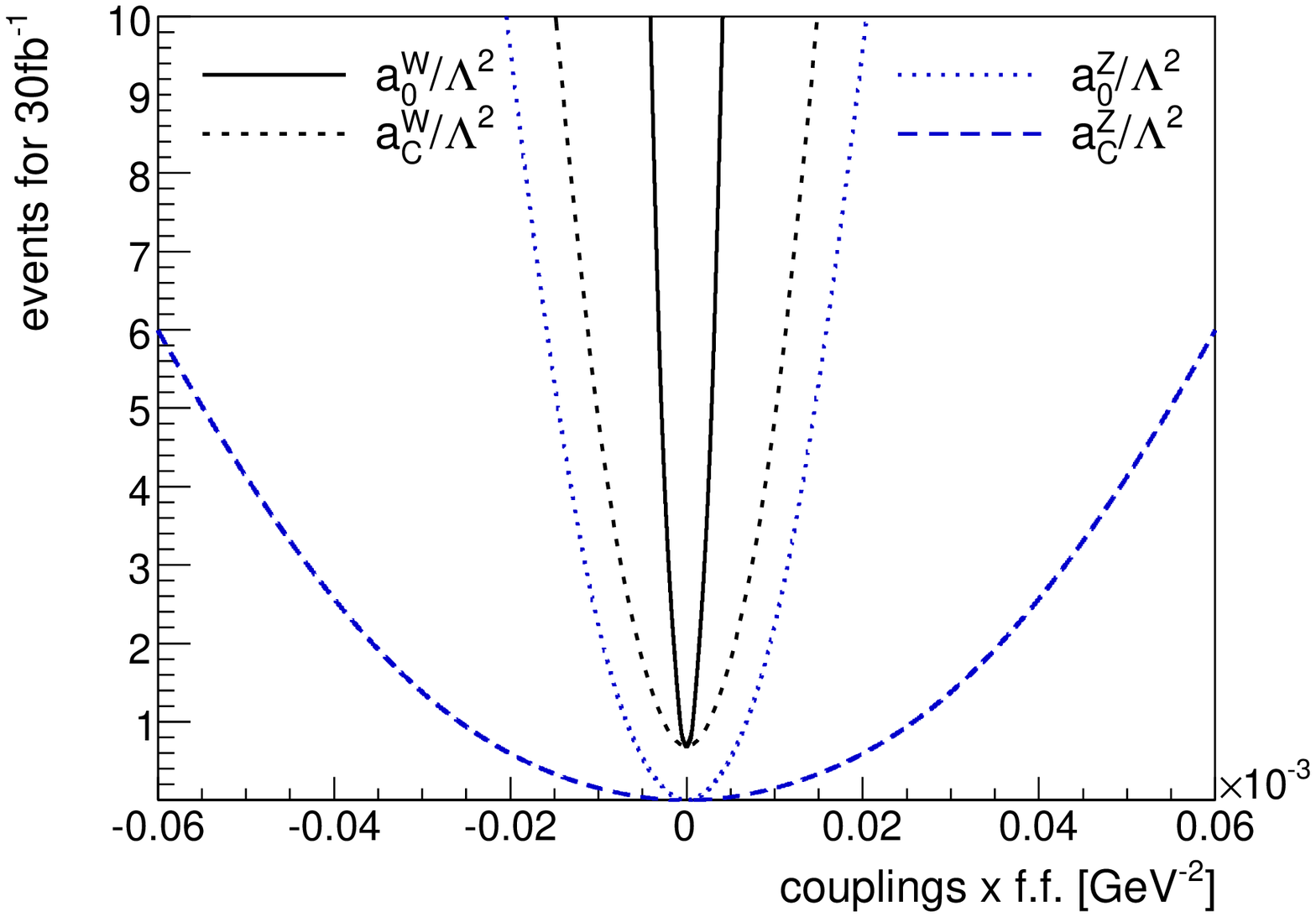,width=6.cm} 
\caption{Number of events for signal due to different values of 
anomalous couplings after all cuts (see text) for a luminosity of
30 fb$^{-1}$.}

\end{minipage}
\hfill
\begin{minipage}[t]{.45\textwidth}

\epsfig{file=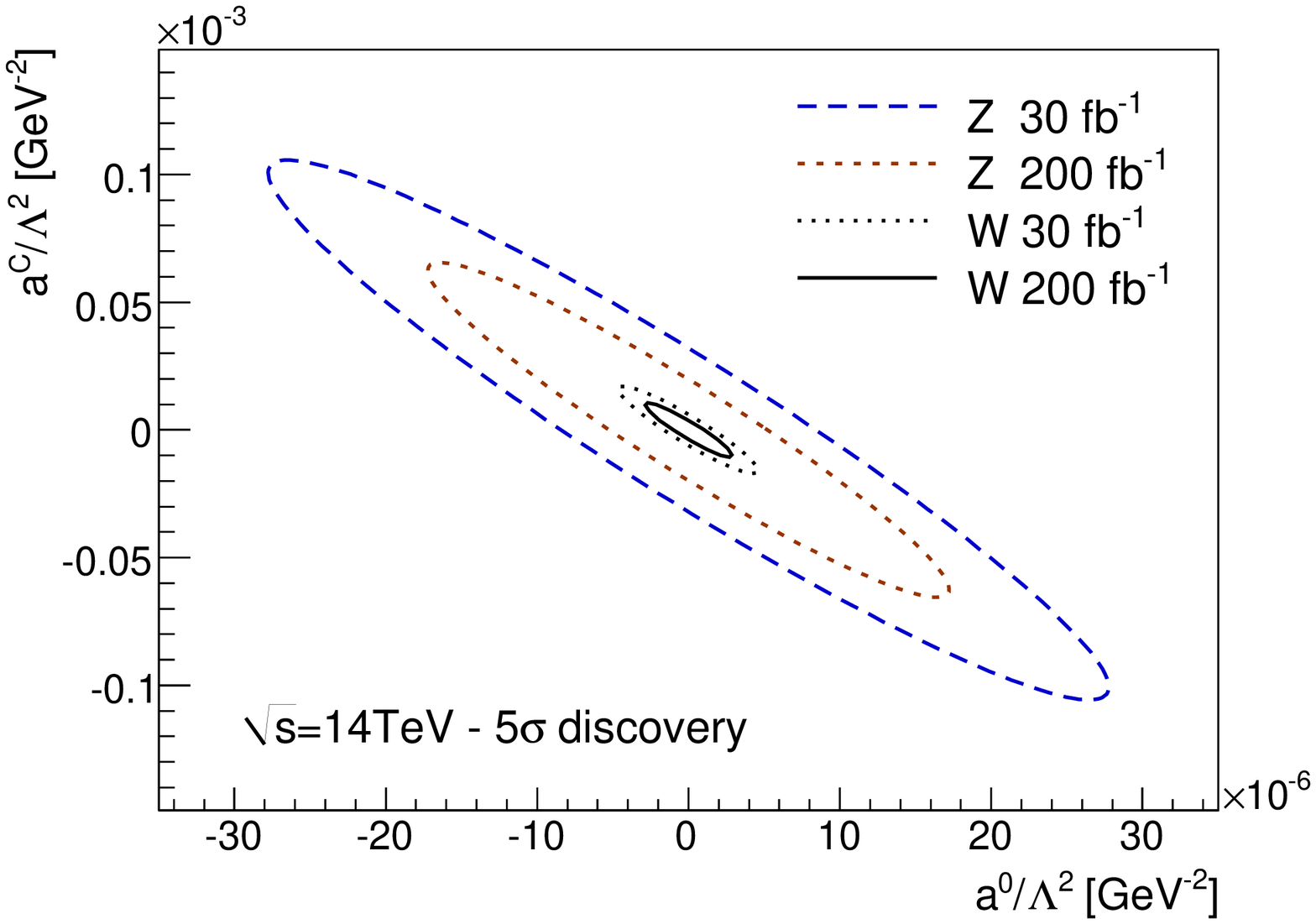,width=6.cm} 
\caption{$5\sigma$ discovery contours for all the $WW$ and $ZZ$ quartic 
couplings at $\sqrt{s}=14$ TeV for a luminosity of 30 fb$^{-1}$ and 200
fb$^{-1}$.}

\end{minipage}
\hfill
\end{figure}

\end{document}